\documentclass[epj]{svjour}
\usepackage{amssymb}
\usepackage{graphics}
\begin{document}
\title{\bfseries DIFFERENTIAL CROSS SECTION OF DP-ELASTIC
SCATTERING AT INTERMEDIATE ENERGIES}
\author{N. B. Ladygina\inst{1} 
\thanks{\emph{Present address:} nladygina@jinr.ru}%
}                     
\offprints{}          
\institute{Joint Institute for Nuclear Research, LHEP, Dubna, Russia}
\date{Received: date / Revised version: date}
%
\abstract{
 The deuteron-proton elastic scattering is studied in the multiple
 scattering expansion formalism. The contributions of the one-nucleon-exchange,
single- and double scattering are taken into account.
The Love and Franey parameterization of the nucleon-nucleon $t$-matrix is used,
that gives an opportunity to include the off-energy-shell 
effects into calculations.
Differential cross sections are considered at four energies, 
$T_d=390, 500, 880, 1200$ MeV.
The obtained results are compared with the experimental data.
\PACS{{21.45.+v}{Few-body systems}\and {25.45.-z}{2H-induced reactions}
\and {25.45.De}{Elastic and inelastic scattering}
\and {24.10.Jv}{Relativistic models}
     } 
} 
\maketitle
\section{Introduction}
\label{intro}
The study of the deuteron-proton elastic scattering has a longtime
story. The first nucleon-deuteron experiments were performed already in fifties
of the previous century \cite{scham}-\cite{postma}. Differential cross sections \cite{scham}-\cite{crewe}
and polarization  \cite{marshal}-\cite{postma}
were measured at few hundred MeV. 
Nowadays this reaction is still the subject of investigations
\cite{cr500}-\cite{pkur}.
This process is the simplest 
example of the hadron nucleus collision that is why the interest to 
this reaction is justified. A number  of experiments 
on deuteron- nucleon elastic scattering is aimed at getting some information
about the deuteron wave function and nucleon-nucleon amplitudes
from $Nd$ scattering observables. Moreover, the study of the reaction 
mechanisms,
investigations of the few-body scattering dynamics are also very 
important to understand the nature of nuclear interactions.

A good theoretical description of the deuteron-nucleon process was 
obtained for low energies, where the multiple scattering 
formalism
based on the solution of the Faddeev equations, has been applied to
solve this
problem   \cite {glrew}. However,  at the energies 
above 150 MeV 
there is some discrepancy 
between the experimental data and theoretical predictions
in the minimum of the differential cross section. At present 
many efforts are undertaken to extend the Faddeev calculation technique
into the relativistic regime \cite{elster72}-\cite{elster78}. But up to now there are no
reasonable descriptions of the experimental data obtained in the
Faddeev equation framework at intermediate energies.

Also p-d scattering at low energies was considered in the approach based on the
solution of the three-particle Schr{\"o}dinger equation using the Kohn
variational principle (KVP)\cite{kiev30},\cite{deltuva65}. Special attention
in these works was given to the study of the Coulomb effects. It has been shown that
at the energies below 30 MeV the influence of the Coulomb interaction is appreciable,
while it  considerably reduces at $T_{lab}=65$ MeV \cite{deltuva65}.

The high-energy deuteron-proton scattering in the forward hemisphere
 is successfully
 described by 
the Glauber theory which takes both single and double interactions into account
 \cite{franco},\cite{francol}. In \cite{har} it is shown that filling of 
the minimum,
due to the interference between the single- and double-scattering amplitudes,
is  explained by the presence of the D-wave in the deuteron wave function.

However, as it is shown in \cite{alberi}, \cite{japh},
the off-energy-shell effects begin to play an important role at scattering
angles larger than $30^0$ in c.m.  
In \cite {kaptari} the
deuteron-proton backward elastic scattering was considered
in the Bethe-Salpeter approach. Here various relativistic
effects   were  studied. 
It has been shown, that the relativistic corrections coming from  
the negative energy P-waves
are negligible while Lorentz boost effects become evident 
already at  $P_{lab}\sim 0.2\div 0.3$ GeV/c
(here,$P_{lab}$ is the scattered proton momentum in the laboratory frame).

The present paper considers the intermediate energy range from 200 MeV
up to 600 MeV of the initial nucleon. 
On the one hand, the energies are not large enough  to apply the
Glauber theory. 
On the other hand, in this region  it is already necessary to take
relativistic effects into account and as a consequence
the Faddeev calculation technique does not work properly at these energies.
Nevertheless, at these energies it is still possible to use
a non-relativistic deuteron wave function due to transfer
into the deuteron Breit frame.
Here, it is also very important to describe correctly the nucleon-
nucleon vertices, namely the spin structure as well as angular and energy dependences.

In the previous paper \cite{japh} the polarization observables, such as
vector and tensor analyzing powers, were studied in the 
presented approach.
In this work only the differential cross sections  are considered
at four energies. The energy range has been chosen to demonstrate the
region, where this model is justified.

Sect.2 gives  the expression for the
differential cross section and
general kinematical definitions.
 The theoretical formalism is presented in Sect.3.
We consider a multiply-scattering series up to the second-order terms
of the nucleon-nucleon $t$-matrix. 
The iterations of the Alt-Grassberger-Sandhas equations is
performed to get this expansion. The wave function of the moving deuteron,
 which depends on the two variables, is used in the calculations.

We apply the $NN$ $t$-matrix constructed by Love and Franey
\cite {LF} in order  to describe the nucleon-nucleon interactions. 
This parameterization allows one to take the energy and angular dependences into 
account. Moreover, it is possible to extend this approach to the off-energy-shell
region. The results of the calculations are
discussed in Sect.4. The conclusions are given in Sect.5.

\section{Kinematics}
\label{kinematics}

A  cross section of the  $dp$-scattering is expressed via the squared reaction amplitude:
\begin{eqnarray}
\label{crsec}
&&\sigma (dp\to dp)=(2\pi)^4\frac{1}{6}\int\frac{d\vec P_d^\prime}{E_d^\prime}\frac{d\vec p_p^\prime}{E_p^\prime}\delta (E_d+E_p-E_d^\prime-E_p^\prime)
\nonumber\\
&&\delta (\vec P_d+\vec p-\vec P_d^\prime-\vec p^\prime)
\frac {|\sqrt{E_d^\prime E_p^\prime}~~ {\cal J} \sqrt{E_d E_p}|^2}{\sqrt{(P_d p)^2-P_d^2 p^2}}.
\end{eqnarray}
This expression is valid for an arbitrary frame. Here $(E_d,P_d)$, $(E_p,p)$ are the energies and momenta of the
initial deuteron and proton, respectively. The corresponding variables for the final particles are defined
with the primed symbols. Starting from Eq.(\ref{crsec}) one can obtain a formula for the differential cross
section in the center-of-mass,  
\begin{equation}
\frac {d\sigma}{d\Omega ^*}=(2\pi)^4\cdot\frac{1}{6}\cdot\frac{1}{s}|\sqrt{E_d^\prime E_p^\prime}~~{\cal J} \sqrt{E_d E_p}|^2.
\end{equation}
It should be emphasized that expression \\
$|\sqrt{E_d^\prime E_p^\prime}~~{\cal J} \sqrt{E_d E_p}|^2$ is invariant.
Since the reaction amplitude ${\cal J}$ is calculated in the Breit frame,
the energies should be also defined in the same system:
\begin{eqnarray}
E_d&=&E_d^\prime=\sqrt{M_d^2+\vec Q^2}, ~~~\vec Q^2=-t/4,~~~
\\
E_p&=&E_p^\prime=\sqrt{m_N^2+\vec p^2},~~~~(\vec p\vec Q)=-\vec Q^2,
\end{eqnarray}
where $M_d$ and $m_N$ correspond to the deuteron and nucleon masses, respectively,
and $\vec Q$, $\vec p$ are momenta of the deuteron and proton in the Breit frame.
The Mandelstam variables $s$ and $t$ are defined through the laboratory deuteron
energy and center-of-mass angle, $\theta^*$, 
\begin{eqnarray}
s&=&(P_d+p)^2=M_d^2+m_N^2+2m_N E_d^{lab}
\nonumber\\
t&=&(P_d-P_d^\prime)^2=-\frac{4m_N^2}{s}(\vec P_d^{lab})^2(1-\cos\theta^*).
\end{eqnarray}

\section{General formalism}
\label{formalism}

 According to the three-body collision theory,
the amplitude of the deuteron-proton elastic scattering $\cal J$
is defined by the matrix element of the transition operator $U_{11}$:
\begin{eqnarray}
U_{dp\to dp}&=&\delta(E_d+E_p-E^\prime_d-E^\prime _p) {\cal J}=
\nonumber\\
&&<1(23)|[1-P_{12}-P_{13}]U_{11}|1(23)>.
\end{eqnarray}
Here, the state $|1(23)>$ corresponds to the configuration, 
when  nucleons 2 and 3
form the deuteron state and nucleon 1 is free. 
Emergence of the permutation operators for two nucleons $P_{ij}$
reflects the fact that the initial and final states are antisymmetric
due to the two particles exchange.

The transition operators for rearrangement scattering 
 are defined by the Alt--Grassberger--Sandhas equations
 \cite{Alt}--\cite{AGS}:
\begin{eqnarray}
U_{11}&=&~~~~~~~~t_2g_0U_{21}+t_3g_0U_{31},
\nonumber\\
U_{21}&=&g_0^{-1}+t_1g_0U_{11}+t_3g_0U_{31},
\\
U_{31}&=&g_0^{-1}+t_1g_0U_{11}+t_2g_0U_{21},
\nonumber
\end{eqnarray}
where $t_1=t(2,3)$, etc., is the $t$-matrix of the two-nucleon interaction and
$g_0$ is the free three-particle propagator. The indices $ij$ for the
transition operators $U_{ij}$ denote free particles $i$ and $j$ in the final
 and initial states, respectively.

Iterating these equations up to the $t_i$-second-order terms,  we can present
the reaction amplitude as a sum of the following three contributions:
 one-nucleon exchange,
single scattering, and double scattering:
\begin{eqnarray}
\label{contrib}
{\cal J}_{dp\to dp}&=&{\cal J}_{\rm ONE}+{\cal J}_{\rm SS}+{\cal J}_{\rm DS},
\nonumber\\
{\cal J}_{\rm ONE}&=&-2<1(23)|P_{12}g_0^{-1}|1(23)>,
\nonumber\\
{\cal J}_{\rm SS}&=&2<1(23)|t_3^{\rm sym}|1(23)>,
\\
{\cal J}_{\rm DS}&=&2<1(23)|t_3^{\rm sym}g_0t_2^{\rm sym}|1(23)>,
\nonumber
\end{eqnarray}
where we have introduced notations for antisymmetrized operators
$t_2^{\rm sym}=[1-P_{13}]t_2$  and $t_3^{\rm sym}=[1-P_{12}]t_3$. 
Schematically, this sequence is shown in Fig.1.

The convergence of the multiple scattering series was
studied in ref.\cite{witala},\cite{elster78}. A reasonable agreement between 
the results of the calculations taking the second order corrections
into account and the full Faddeev calculations, was obtained at the
laboratory energies up to 2 GeV at least for the forward scattering angles.
The analogous conclusion can be done for the differential cross section
of the elastic scattering at $E_{lab}=400$ MeV for the scattering
angle up to $140^0$ \cite{witala}. It gives us the ground to 
suppose that the reduced series (\ref{contrib}) is a good approximation
to describe dp-elastic scattering at the energies above 400 MeV.

\begin{figure*}
\hspace*{1cm}
\resizebox{0.6\textwidth}{!}{
\includegraphics{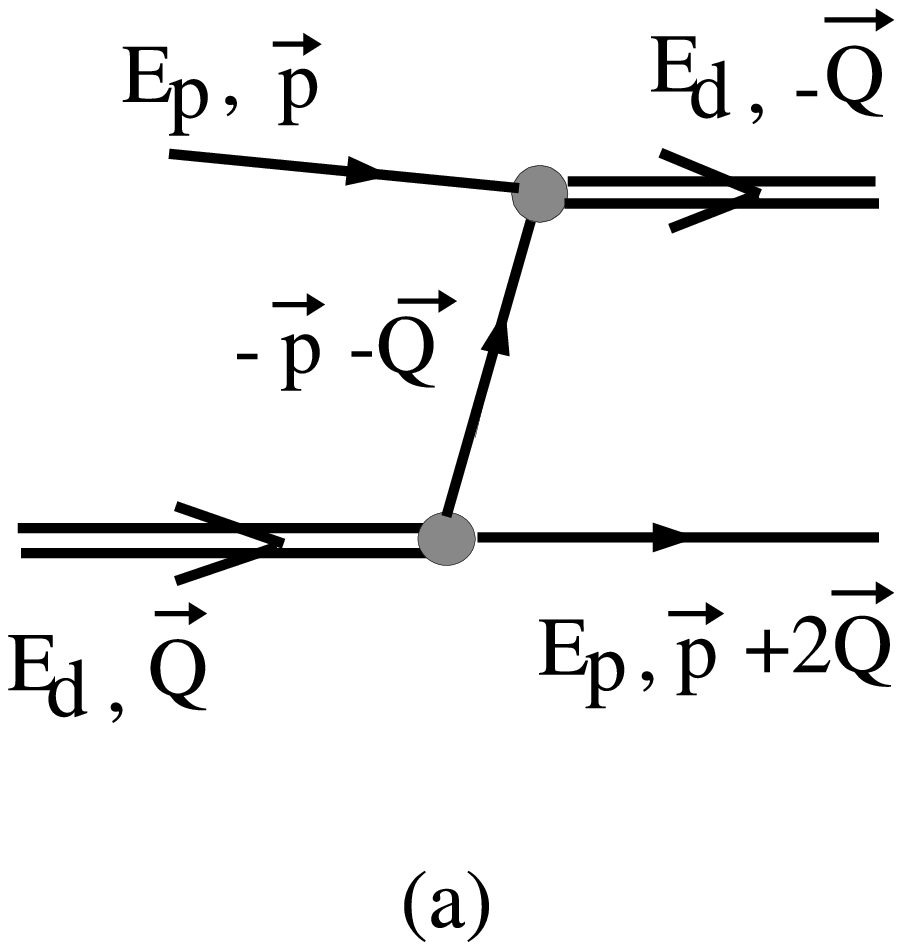}
}
\hspace*{-1cm}
\resizebox{0.6\textwidth}{!}{
\includegraphics{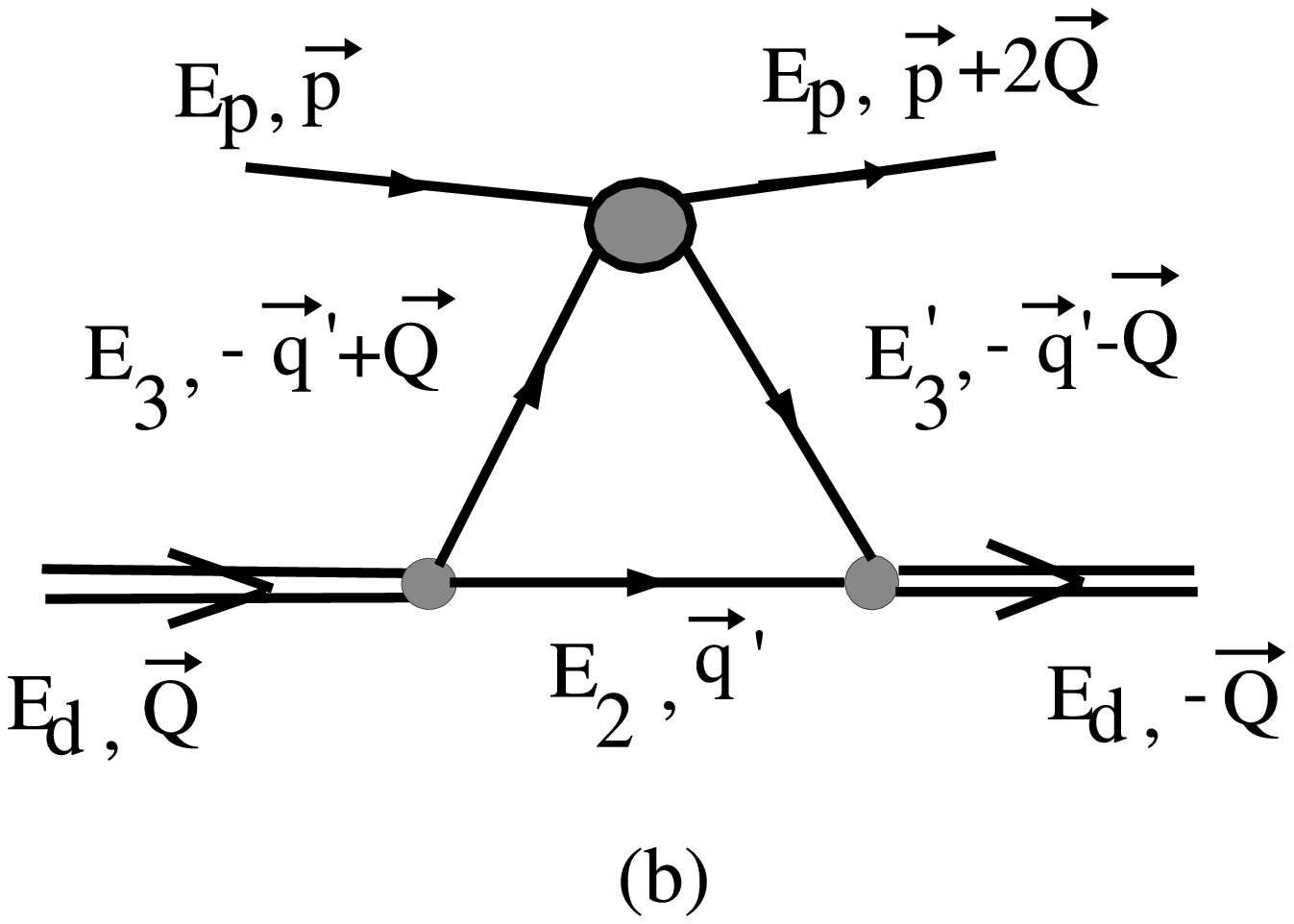}
}

\vspace*{-8cm}
\hspace*{4cm}
\resizebox{0.6\textwidth}{!}{
\includegraphics{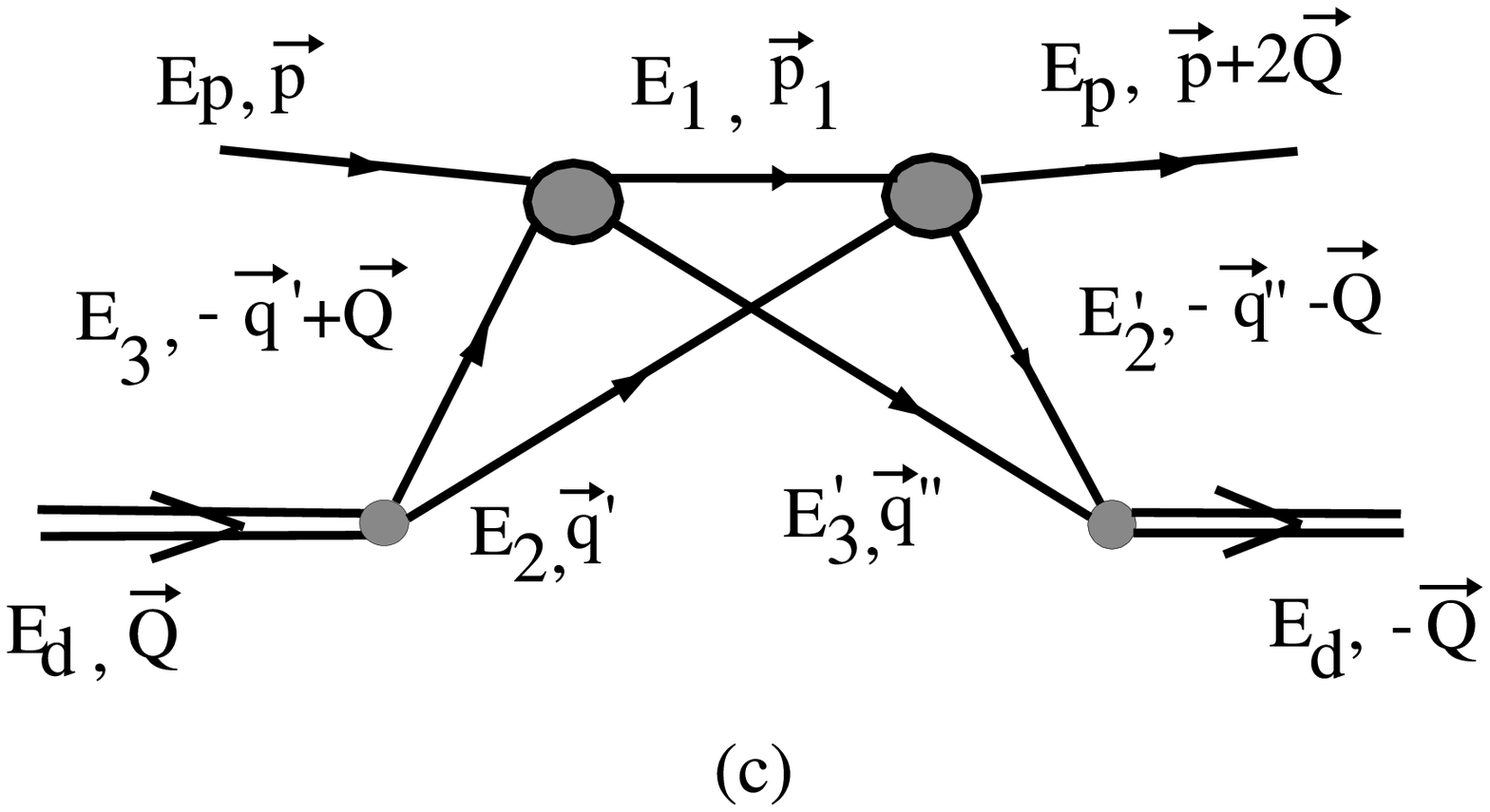}
}
\vspace*{-7cm}
\caption{
The diagrams are taken into consideration: one-nucleon-exchange (a), single
scattering (b), and  double scattering (c) graphs.}
\label{fig:1}
\end{figure*}

After straightforward calculations we have got the following expressions:
\\
{\it for the one-nucleon-exchange amplitude --}
\begin{eqnarray}
{\cal J}_{ONE}&=&-\frac{1}{2}(E_d-E_p-\sqrt{m_N^2+\vec p~^2-\vec Q\-^2})
\cdot
\\
&&\hspace{-1.5cm}
<\vec p^\prime \mu^\prime;-\vec Q {\cal M}_d^\prime|
\Omega^\dagger_d(23)
[1+(\mbox{\boldmath$\sigma_1\sigma_2$})]
\Omega_d(23)|
\vec Q {\cal M}_d;\vec p \mu >~~,
\nonumber
\end{eqnarray}
\\
{\it for the  single scattering amplitude --}
\begin{eqnarray}
\label{ss}
{\cal J}_{SS}&=&\int d\vec q~^\prime <-\vec Q {\cal M}_d^\prime |
\Omega_d^\dagger|
\vec q~^\prime \mu_2, -\vec Q-\vec q~^\prime \mu_3^\prime>
\\
&&\hspace{-1cm}
<\vec p~^\prime \mu^\prime, -\vec Q-\vec q~^\prime \mu_3^\prime |
\frac{3}{2}t^1_{13}(E)+\frac{1}{2}t^0_{13}(E)|
\vec p \mu, \vec Q -\vec q~^\prime \mu_3>
\nonumber
\\
&&\hspace{-1cm}
<\vec q~^\prime \mu_2,\vec Q-\vec q~^\prime \mu_3|
\Omega_d|\vec Q {\cal M}_d>~~.
\nonumber
\end{eqnarray}
\\
{\it for the double scattering amplitude --}
\begin{eqnarray}
\label{ds}
&&{\cal J}_{DS}=\int d\vec q~ ^\prime\int d\vec q^{\prime\prime }
<-\vec Q {\cal M}_d^\prime|\Omega_d^\dagger 
\\
&&|-\vec Q-\vec q~^{\prime\prime}~\mu_2^\prime,
\vec q~^{\prime\prime}~ \mu_3^\prime>
<\vec p^\prime~ \mu^\prime,
 -\vec Q-\vec q~^{\prime\prime}~\mu_2^\prime, \vec q~^{\prime\prime}~\mu_3^\prime|
\nonumber\\
&&\frac{ 
t_{12}^1(E^\prime) 
t_{13}^1(E)+
[t_{12}^1(E^\prime)+
t_{12}^0(E^\prime)]
[t_{13}^1(E)+
t_{13}^0(E)]/4}{E_d+E_p-E_1 -E_2-E_3^\prime +i\varepsilon}
\nonumber\\
&&|\vec p~\mu, \vec q~^\prime~ \mu_2, \vec Q -\vec q~^\prime~\mu_3>
\nonumber\\
&&
<\vec q~^\prime~ \mu_2,~\vec Q -\vec q~^\prime~ \mu_3|\Omega_d|\vec Q {\cal M}_d>.
\nonumber
\end{eqnarray}
Here the spin   projections are denoted as
$\mu$ for nucleons and ${\cal M}_d$ -- for deuterons.
Henceforth, all summations over dummy discrete indices are implied.
The superscript of the $t$-matrix corresponds to the isotopic spin 
of the nucleon-nucleon state. The argument of the $NN$-matrix is defined
as the three-nucleon on-shell energy excluding the energy of the nucleon which does not
participate in the interaction:
\begin{eqnarray}
E=E_d+E_p-E_2,~~~~~~E^\prime=E_d+E_p-E^\prime _3.
\end{eqnarray}
The three-nucleon free propagator in Eq.(\ref {ds}) can be decomposed on two
terms using the well-known formula: 
\begin{eqnarray}
&&\frac{1}{E_d+E_p-E_1 -E_2 -E_3^\prime +i\varepsilon}=
\\
&&{\cal P}\frac{1}{E_d+E_p-E_1 -E_2 -E_3^\prime }
\nonumber\\
&&
-i\pi\delta(E_d+E_p-E_1 -E_2 -E_3^\prime ).
\nonumber
\end{eqnarray}
 
The principal value part is often neglected to simplify the further calculations.
However, as it was shown in \cite{japh}, it is very important 
to take into account the full representation of the
three-nucleon propagator, especially, at the scattering angles above $40^0$.

All the calculations are performed in the deuteron Breit frame, where
the deuterons move  in opposite directions with equal momenta (Fig.1).
It allows us to minimize  the relative momenta of the nucleons in the both
deuterons. As a consequence, the non-relativistic deuteron wave function
can be applied in the energy range under consideration.

In the rest frame the non-relativistic wave function  of the deuteron 
depends only on one variable $\vec p_0$, which is the
 relative momentum of the  outgoing proton and neutron:
\begin{eqnarray}
\label{dwf0}
&&<\mu_p \mu_n|\Omega_d|{\cal M}_d>=
\frac{1}{\sqrt{4\pi}}<\mu_p \mu_n|\{ u(p_0)+
\\
&&\hspace*{2cm}
\frac{w(p_0)}{\sqrt 8}
[3(\mbox{\boldmath$\sigma_1$} \hat p_0)(\mbox{\boldmath$\sigma_2$}\hat p_0)-(
\mbox{\boldmath $\sigma_1$ $\sigma_2$}
)]
\}|{\cal M}_d>,
\nonumber
\end{eqnarray}
where $u(p_0)$ and $w(p_0)$ describe the $S$ and $D$ components of 
the deuteron wave function   \cite{B}, \cite{cd}, \cite{par}, $\hat p_0$ is 
the unit vector in $\vec p_0$ direction.

In order to get the wave function of the moving deuteron, it is necessary to apply
the Lorenz transformations for the kinematical variables and Wigner rotations for 
the spin states. This procedure has been expounded in \cite{japh}, and, here,
we give only the result:
\begin{eqnarray}
&&<\vec p_1~\mu_1, \vec p_2 \mu_2|\Omega_d |\vec Q,{\cal M}_d>=
<\vec p_1~\mu_1, \vec p_2 \mu_2|g_1(\vec k,\vec Q)+
\nonumber\\
&&g_2(\vec k,\vec Q)(\mbox{\boldmath$\sigma_1$}\vec n)
(\mbox{\boldmath$\sigma_2$}\vec n)+
g_3(\vec k,\vec Q)(\mbox{\boldmath$\sigma_1 \sigma_2$})+
\\
&+&g_4(\vec k,\vec Q)(\mbox{\boldmath$\sigma_1$}\hat k)
(\mbox{\boldmath$\sigma_2$}\hat k)+
g_5(\vec k,\vec Q)((\mbox{\boldmath$\sigma_1$}+
\mbox{\boldmath$\sigma_2$})\vec n)+
\nonumber\\
&+&g_6(\vec k,\vec Q)[(\mbox{\boldmath$\sigma_1$}\hat k)
(\mbox{\boldmath$\sigma_2$},\vec n\times\hat k)+
(\mbox{\boldmath$\sigma_1$},\vec n\times\hat k)
(\mbox{\boldmath$\sigma_2$}\hat k)]|\vec Q,{\cal M}_d>.
\nonumber
\end{eqnarray}
Note, 
the wave function of the moving deuteron is the function of two variables:
the deuteron momentum $\vec Q$ and neutron-proton relative momentum $\vec k$ --
\begin{eqnarray}
\label{mom}
&&\vec Q=\vec p_1 +\vec p_2,
\nonumber\\
&&\vec k=\frac{(E_2+\sqrt {s}/2)\vec p_1-(E_1+\sqrt{s}/2)\vec p_2}{E_1+E_2+\sqrt{s}}.
\end{eqnarray}
Here $(E_1,\vec p_1)$ and $(E_2,\vec p_2)$ are the energy and momentum of the
neutron and proton in the deuteron, and 
 $s$ is the squared sum of the neutron and proton 4-momenta, $s=(p_1+p_2)^2$.
The normal to $(\vec p_1, \vec p_2)$-plane is denoted as $\vec n$.

In general, functions $g_i$ can be obtained by solving a relativistic equation,
 as it was done in \cite {karmanov1}--\cite {carbonell} in the light front model.
But in the present paper a usual non-relativistic DWF is taken as an input.
Therefore, functions $g_i$ are defined as the linear combinations of $u$ and
$w$ ($S$- and $D$-waves). 

In order to describe the nucleon- nucleon  interaction in a wide energy
region, we have used the Love and Franey parameterization \cite {LF} of the
 $NN$ $t$-matrix defined in the center-of-mass as
\begin{eqnarray}
\label{tnn}
&&<\mbox{\boldmath$\kappa$}^\prime  \mu_1^\prime \mu_2^\prime |t_{c.m.}|
\mbox{\boldmath$\kappa$} \mu_1\mu_2>
=<\mbox{\boldmath$\kappa$}^\prime  \mu_1^\prime \mu_2^\prime |
A+
\\
&&B(\mbox{\boldmath$\sigma_1$} \hat N^*)(\mbox{\boldmath$\sigma_2$} \hat N^*)+
C(\mbox{\boldmath$\sigma_1$} +\mbox{\boldmath$\sigma_2$} )\cdot \hat N^* +
D(\mbox{\boldmath$\sigma_1$} \hat q^*)(\mbox{\boldmath$\sigma_2$} \hat q^*) +
\nonumber\\
&&
F(\mbox{\boldmath$\sigma_1$} \hat Q^*)(\mbox{\boldmath$\sigma_2$} \hat Q^*)
|\mbox{\boldmath$\kappa$} \mu_1\mu_2>.
\nonumber
\end{eqnarray}
Here the orthonormal basis is  combinations of the nucleons'
relative momenta in the initial \mbox{\boldmath$\kappa $} and final 
\mbox{\boldmath$\kappa$}$^\prime$
states:
\begin{equation}
\hat q^*=\frac {\mbox{\boldmath $\kappa$} -\mbox{\boldmath$\kappa$}^\prime}
{|\mbox{\boldmath$\kappa$} -\mbox{\boldmath$\kappa$}^\prime|},~~
\hat Q^*=\frac {\mbox{\boldmath$\kappa$} +\mbox{\boldmath$\kappa$}^\prime }
{|\mbox{\boldmath$\kappa$} +\mbox{\boldmath$\kappa$}^\prime|},~~
\hat N^*=\frac {\mbox{\boldmath$\kappa$}  \times 
\mbox{\boldmath$\kappa$}^\prime }{|\mbox{\boldmath$\kappa$}
\times\mbox{\boldmath$\kappa$}^\prime |}.
\end{equation}
The amplitudes $A,B,C,D,F$ are the functions from the center-of-mass
energy and scattering angle. The radial parts of these amplitudes are taken 
as a sum of Yukawa terms. A new fit of the model parameters was done \cite{newlf} 
according to the current phase-shift-analysis data SP07 \cite{said}.

Since the matrix elements are expressed
through the effective $NN$-interaction operators sandwiched
between the initial and final plane-wave states, this construction 
can be extended to the off-shell case allowing the initial and final
states to get the current values of the \mbox{\boldmath $\kappa$} and
\mbox{\boldmath $\kappa^\prime$}. Obviously, this extrapolation does 
not change the general spin structure.

To use  definition (\ref{tnn}) for the nucleon-nucleon matrix in the further calculation, 
it is necessary to relate 
 $t$-matrices in the two-nucleons c.m. and  in the
deuteron  Breit frame. This relation can be obtained by using the Lorenz transformation
and Wigner rotation technique as it was done for the deuteron wave function.
This problem has been considered in detail in \cite{fbs}, \cite{japh}.
\begin{eqnarray}
\label{tlab}
&&<\vec p^\prime \vec p_3^\prime; \mu^\prime \mu_3^\prime|t(E)|
\vec p \vec p_3; \mu \mu_3>=
{\cal N F}<\kappa^\prime  \mu_1^\prime \mu_2^\prime |
\nonumber\\&&
W^\dagger_{1/2 } (\vec p^\prime )
W^\dagger_{1/2 }(\vec p_3^\prime )
 t_{c.m.}(\sqrt{s})~ W_{1/2} (\vec p)
W_{1/2 }(\vec p_3)|
\kappa \mu_1\mu_2>.
\nonumber\\
\end{eqnarray}
Here, momentum notations correspond to the ones given in Fig.1b.
The Wigner rotations of the initial and final states are performed
around $\vec n$ and $\vec n^\prime$  axes, respectively:
\begin{eqnarray}
\vec n=\frac{\vec p\times \vec p_3}{|\vec p\times \vec p_3|},~~~~~~~~
\vec n^\prime=\frac{\vec p~^\prime\times \vec p_3~^\prime }
{|\vec p~^\prime \times \vec p_3~^\prime |}.
\end{eqnarray}
The Wigner rotation operator in the spin space of the $i$-th nucleon
has the standard form:
\begin{eqnarray}
\label{w12}
&&W_{1/2}(\vec p_i, \vec u)={\rm exp}\left\{ -i\omega_i(\vec n_i
\mbox{\boldmath$\sigma_i$})/2\right\} =
\\
&&{\rm cos}(\omega_i/2)[1-i(\vec n_i\mbox{\boldmath$\sigma_i$}){\rm tg}
(\omega_i/2)],
\nonumber
\end{eqnarray}
where $\omega_i$ is the angle of this rotation. 
The normalization factor ${\cal N}$ provides conditions of orthonormality and completeness
  and  is defined by the Jacobian of the transformations. The kinematical factor 
${\cal F}$ is due to the relation between two off-energy shell $t$-matrices, one of them depends
on zero total momentum   \cite{garcilazo}, \cite{japh}.

The argument of the  $t$-matrix in the right-hand side of Eq.(\ref {tlab}) is
the square root of the 
Mandelstam variable $s$, which is defined through the kinematical variables in the
deuteron Breit  frame taken  on the mass-shell:
\begin{eqnarray}
s=E^2-\vec K^2,
\end{eqnarray}
where $\vec K=\vec p+\vec p_3$ is the total momentum of the nucleon-nucleon pair in 
the frame of the calculation.

\begin{figure}
\resizebox{0.54\textwidth}{!}{
  \includegraphics{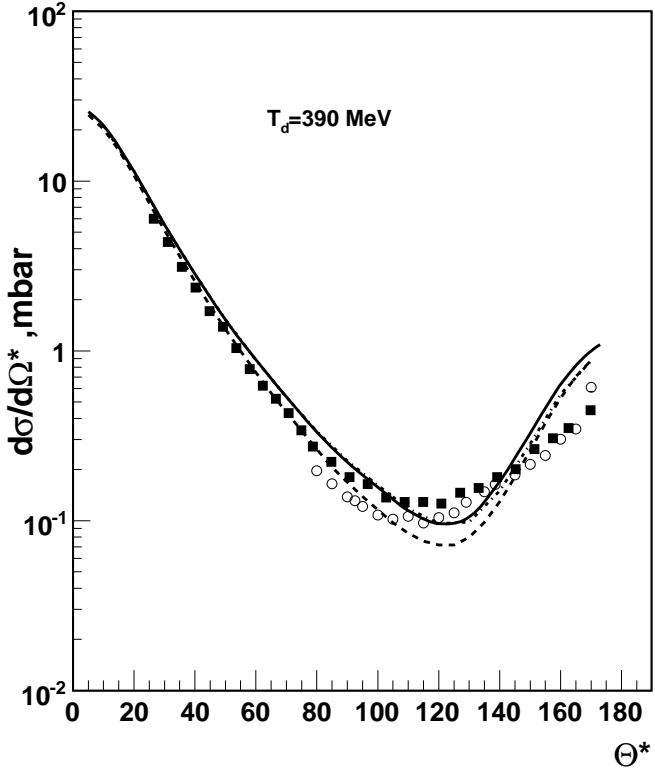}
}
\caption{The differential cross section at the deuteron kinetic 
energy of 390 MeV as a function of the c.m. scattering angle.
The dashed line corresponds to the calculations including only
one-nucleon-exchange and single scattering diagrams
into consideration. The solid
and dashed-dotted lines are the results of the calculations
taking into account also the double scattering diagram with the principal
value part of the free propagator and without it, respectively. 
The data are taken from \cite{ermish} ($\blacksquare$) and \cite{adelberger}($\bigcirc$).}
\label{fig:2}       
\end{figure}

\begin{figure}
\resizebox{0.54\textwidth}{!}{
  \includegraphics{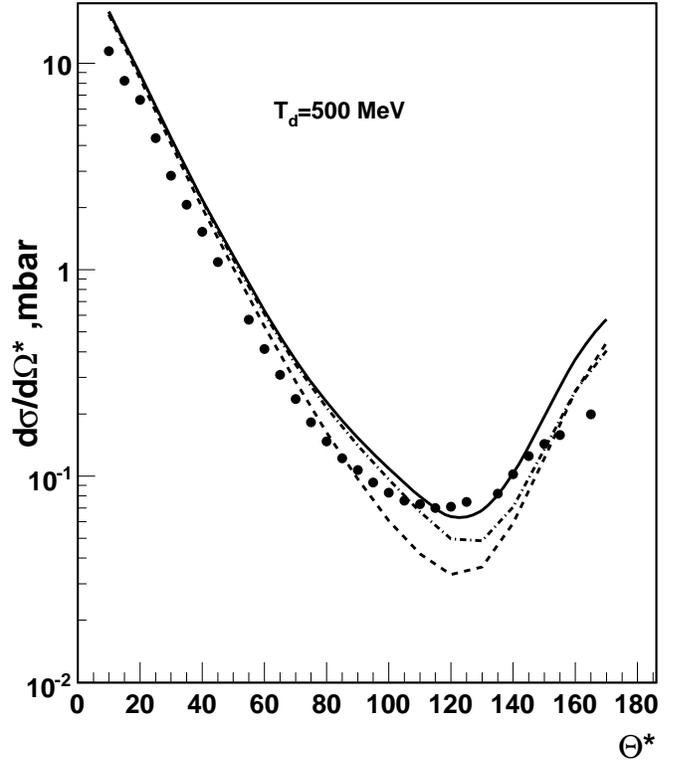}
}
\caption{The differential cross section at the deuteron kinetic 
energy of 500 MeV as a function of the c.m. scattering angle. 
The lines are the same as in Fig.\ref{fig:2}.
The data are taken from\cite{cr500}.}
\label{fig:3}      
\end{figure}

\section{Results and discussion}
\label{result}

\begin{figure}
\resizebox{0.54\textwidth}{!}{
  \includegraphics{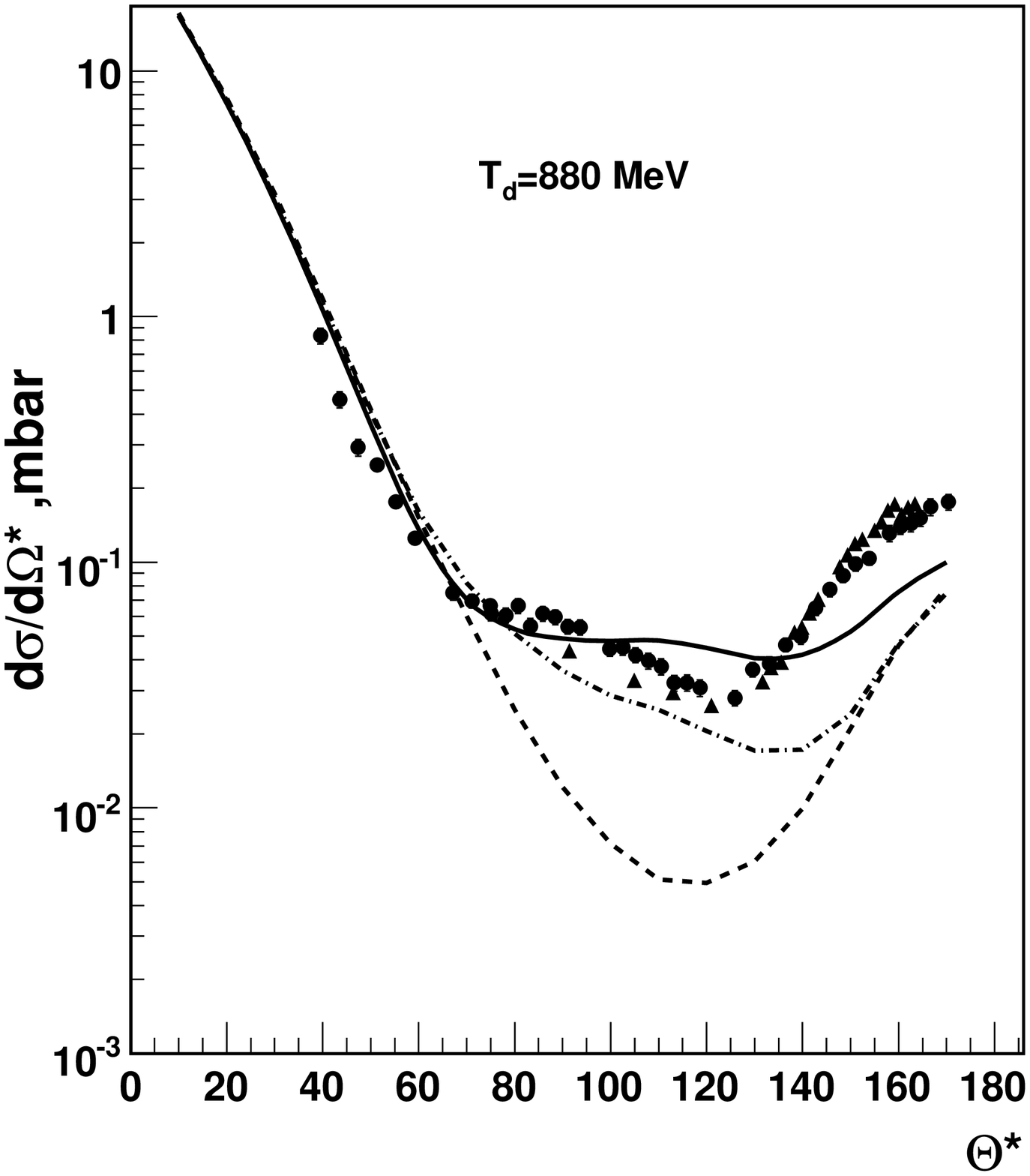}
}
\caption{The differential cross section at the deuteron kinetic 
energy of 880 MeV as a function of the c.m. scattering angle.
The lines are the same as in Fig.\ref{fig:2}.  
The data are taken from \cite{cr880_a} ($\bullet$), and\cite{cr880_b}($\blacktriangle$).}
\label{fig:4}      
\end{figure}

\begin{figure}
\resizebox{0.54\textwidth}{!}{
  \includegraphics{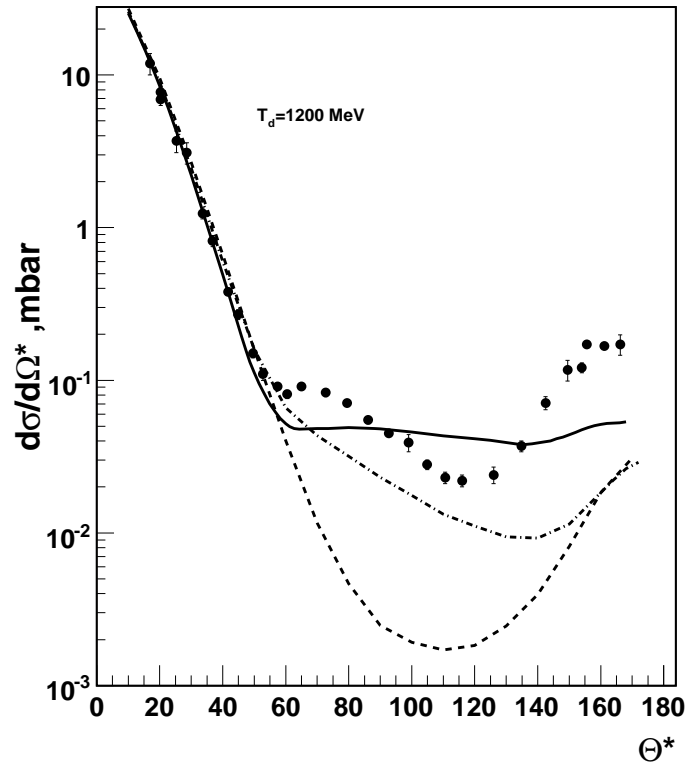}
}
\caption{The differential cross section at the deuteron kinetic 
energy of 1200 MeV as a function of the c.m. scattering angle.
The lines are the same as in Fig.\ref{fig:2}.  
The data are taken from\cite{cr1200}.}
\label{fig:5}       
\end{figure}

The results of the calculations for the differential cross sections
are presented in Figs.\ref{fig:2}-\ref{fig:5}. The four deuteron kinetic 
energies, 390, 500, 880 and 1200 MeV, have been considered.
 All calculations were performed with the CD-Bonn deuteron wave function \cite{cd}.
The dashed line in Figs.\ref{fig:2}-\ref{fig:5} corresponds to 
the calculations taking into account only
ONE and single-scattering contributions. The dash-dotted line represents
 the calculation
including the double-scattering term without the principal value part in the
three-nucleon propagator, and
 the solid line stands for the calculations with the full three-nucleon
propagator.

One can see that the results taking the double scattering  into account
 and without it, are  close to each other up to 50-60 degrees of the scattering angle
for all the considered energies. At these angles the single-scattering gives
the main contribution to the differential cross section. Then the difference
 between these results increases.
It should be noted that the double scattering contribution is not large enough
for the deuteron energy of 390 MeV (Fig.2). As a consequence, the curves
corresponding to the results with and without the principal value part of the
three-nucleon propagator, are practically undistinguished. However, the
double-scattering  effect increases with the energy. 
Already for the energy of 500 MeV (Fig.3) the difference between the results 
taking DS into consideration and without it, is significant -- it is about
2 times in minimum of the differential cross section. For higher energies
this factor is even larger: it is about 10 times for $T_d=880$ MeV (Fig.4), and
about 20 times for the energy equal to 1200 MeV (Fig.5).

The difference between the results taking into account the principal 
value part of the three-nucleon propagator ("full" calculation) and without
it, also increases with the energy. At small angles
this deviation is practically invisible, while it is significant for
the scattering in the backward hemisphere, $\theta^*\ge 90^0$, where the
double scattering effect plays the main role.  Such a discrepancy can be, probably,
explained by the relativistic effects related with the off-energy-shell
behaviour of the nucleon-nucleon $t$-matrices, and Fermi motion of the
 nucleons in the deuterons.  It is interesting to emphasize that the curves
corresponding to the  calculations without the principal value part of the 
free propagator are close to the ONE+SS-results
at the scattering angle above $160^0$ at all the energies under consideration.

The deviation of the theoretical predictions from the experimental data
are observed at the scattering angles above $140^0$. One can assume some reasons of this
distinction. It is possible that the multiple scattering series is not yet converged
in the backward direction, and it is necessary to take the more orders of this series into
account. This is supposed  according to the results of the
investigation performed in ref.\cite{witala}.

 The agreement between the experimental data and theoretical 
predictions could be also improved, if the three-nucleon forces are included into
consideration.
In fact, a good description of the
differential cross sections, both in the diffraction minimum and  the backward direction,
 was obtained at the energies below 400 MeV \cite{wit3N}, when  
the three-nucleon forces were taken into account. Unfortunately, 
at present there are no
such calculations at higher energies.
However, it should be noted, that the three-nucleon forces can be 
effectively included
through the $\Delta$-isobar excitation.
The employment of this approach \cite{deltuva} gives the results, which are in agreement
with the results obtained in \cite{wit3N}.
The essential contribution of the $\Delta$-isobar excitation  into the  differential cross section
was also demonstrated in ref.\cite{kapdelta},
where  the pd-backward elastic scattering was
studied  at the high energies.

Nevertheless, the results of the "full" 
calculations  are in a reasonable agreement with the experimental data up to
the scattering angle of $140^0$ at all the considered energies.   

\section{Conclusion}
\label{conclusion}

In the  paper we have presented a method to calculate 
the amplitude of the deuteron--proton elastic scattering at intermediate
energies. Special attention was given to the questions  connected
with the relativistic effects. The transformation of the deuteron
wave function in the rest frame to a moving system was performed, that
 allowed us to use the non-relativistic DWF
  at rather  high energies.
In order to describe nucleon--nucleon interactions in a wide energy range,
we have used
 parameterization of the $NN$ $t$-matrix. The spin transformation
technique has been also applied to relate this $t$-matrix given in the c.m.
to that in the reference frame.

Using this method  we have managed to describe the experimental
data on the differential cross section at
four energies: 395, 500, 880, and 1200 MeV.
 Good agreements have been
obtained  between the experimental data and the theoretical model calculations
 taking  into consideration the one-nucleon-exchange, single- and 
 double-scattering for all the four energies up to the scattering angle of
$140^0$. The distinctions between the data and theoretical predictions at 
the backward angles should be studied more precisely, that is the subject
for further investigations.

\begin{acknowledgement}
The author is grateful to Dr. V.P. Ladygin for fruitful discussions.
This work has been supported by the Russian Foundation for Basic Research
under grant  $N^{\underline 0}$  07-02-00102a. 
\end{acknowledgement}

\end{document}